\documentclass[12pt]{article}
\usepackage{amsmath}
\usepackage[pdftex]{graphicx}
\usepackage{psfrag}
\usepackage{epsf}
\usepackage{enumerate}
\usepackage{natbib}
\usepackage{verbatim}
\usepackage{url} 
\usepackage{authblk}

 \usepackage{tabularx}
 \usepackage{booktabs}
 \usepackage{hyperref}
 \usepackage{caption} 
 \usepackage{array}   
\usepackage{pdflscape}

\usepackage{rotating}

\pdfoutput=1

\begin{document}

\def\spacingset#1{\renewcommand{\baselinestretch}%
{#1}\small\normalsize} \spacingset{1}


\newcommand{\anon}{1}

\title{\bf Assumptions, Modeling and Risks: Applying ICH M15 to Statistical Modeling \ifnum\anon=1\thanks{On behalf of the EFSPI Statistical Methodology Leaders Group}\fi}
\ifnum\anon=1
  \author[1]{Tobias Mielke \thanks{Corresponding author: tmielke1@its.jnj.com}}
  \author[2]{Mouna Akacha}
  \author[3]{Christian Stock}
  \author[4]{Marc Vandemeulebroecke}
  \author[5]{Cornelia Ursula Kunz}
  \author[6]{Fei Chen}
  \author[2]{Simon Newsome} 
  
  \affil[1]{Johnson and Johnson, Germany}
  \affil[2]{Novartis AG, Switzerland}
  \affil[3]{Boehringer Ingelheim Pharma GmbH \& Co.\ KG, Germany}
  \affil[4]{Bayer BCC AG, Switzerland}
  \affil[5]{Boehringer Ingelheim International GmbH, Germany}
  \affil[6]{Johnson and Johnson, Raritan, NJ, USA}
\else
  \author{Authors Withheld for Blind Review}
\fi

\maketitle

\begin{abstract}
In many instances in phase II and III clinical trials, critical assumptions underlying statistical analyses require discussion with regulators and other stakeholders. Such discussions benefit from a guiding process to promote consistency and clarity, but no widely and consistently applied framework currently exists. In January 2026, the International Council for Harmonisation (ICH) released the ICH M15 Guideline on General Principles for Model-Informed Drug Development, which proposes a systematic framework for evaluating mechanistic models to support regulatory and development decisions. Although it focuses on mechanistic modeling, its structured assessment of model risks and benefits may be equally applicable to other approaches typically requiring a thorough appraisal of methodology and alignment among stakeholders, such as external data use and parametric longitudinal data analysis, and novel methodological approaches more generally. We illustrate and discuss how systematic risk–benefit evaluation of statistical models under the ICH M15 framework could work and propose its use to support acceptance of context-specific modeling strategies and to enhance decision-making transparency.
\end{abstract}

\noindent%
{\it Keywords: Statistical modeling, Assumptions, Model-Informed-Drug-Development}  
\vfill

\newpage
\spacingset{1.45} 
\section{Introduction}
\label{sec1}
Drug development is a costly process with high probability of failure \citep{Sertkaya24}. It is based on a thorough understanding of diseases, combined with scientific innovation to identify new targets and approaches to ultimately act on the disease course. Thorough understanding of the disease does typically not mean “complete understanding”. There are many known and unknown factors, which may impact disease and eventually the treatment outcome in various ways. The decision to develop a drug is based on hypotheses, which are motivated by novel insights into mechanistic pathways or a compound's activity on those pathways. Development may fail due to various reasons, including in particular that the drug is not acting sufficiently on the intended pathway, the hypothesized pathway is not relevant to treating the disease, or endpoints considered are not sensitive to detect treatment benefits.

A key role of statistics in drug development is to support decision-making by quantifying and communicating risks \citep{ChuangSteinKirby2021}. Appropriate decision-making requires understanding of uncertainties and limitations in data, be it historical data, or newly generated data, which is collected in properly designed and executed experiments. The proper design itself requires understanding of available data to inform careful considerations on acceptable risks. For example, Assurance is a helpful metric to complement confirmatory trial designs discussions, which are frequently based on powering using clinically relevant, yet possibly overly optimistic treatment effects \citep{OHagan2005Assurance}. Another example is the estimation of trial duration for time-to-event trials in presence of uncertainty on the control group event rates. Overestimation of the control group event rates may result in excessively long clinical trials, which may pose a significant risk to the value proposition of drugs \citep{JennisonTurnbull1999}. 
Lastly, in dose-finding trials, the location of the $ED_{50}$ (Dose which reaches $50\%$ of the maximum effect) is of large importance for designing informative trials \citep{bretz2005combining, bornkamp2009mcpmod}. Overestimation of the $ED_{50}$ in the design stage of a dose-finding trial may result in trial designs, which allocate too many patients to high doses, thereby not allowing to estimate the dose-response relation. These are just some few of the many examples highlighting the importance of assumptions in the design phase of clinical trials. 

Modeling plays a key role in defining assumptions and hypotheses, which are assessed in pharmaceutical research using well-designed experiments and appropriate statistical methods. The gold standard for testing hypotheses is the use of randomized controlled trials (RCTs). Randomization asserts that uncontrollable, yet possibly influential factors are identically distributed across the study arms in a probabilistic sense. This allows for an unbiased assessment of the impact of the controlled variables, such as treatment or dose \citep{ICHE9}. Conduct of RCTs may be infeasible in certain situations, including rare indications \citep{FDARareDiseases,ICHE8R1} and they may be inefficient in situations with large existing understanding of the control-group response \citep{Viele2014HistoricalControls}. Additionally, RCTs may suffer from intercurrent events, which might distort randomization and thereby complicate interpretation of study results \citep{ich2019e9}. Modeling approaches help in some of these situations to guide or support decisions from an RCT, by utilizing assumptions, which are derived from the scientific understanding of the disease and/or previously conducted clinical trials using mechanistic models \citep{Verma2023QSP}, real-world evidence methodology \citep{FDA2018RWE,Izem2023RWE} or Bayesian approaches \citep{Ruberg2023Bayesian,FDA2026Bayesian}. While addressing data gaps in situations where RCTs may be infeasible or inefficient, the use of external data and parametric models is regarded with caution, as biased analyses may result, if external data and/or a model do not fit the true (generally unknown) data-generating mechanism \citep{Freedman2005Causation,Pocock1976}. \citet{burger2021external} summarize promises with the use of external controls, but also provide an extensive list of potential biases resulting from the use of external control data. Ignoring data external to a trial reduces potential for bias, helps control the type-1 error and it simplifies analyses and interpretation. Yet, it may result in missing important additional information, which may be present in the external data or justified expert's understanding. As such, a decision to disregard a well justified model without impact assessment, due to diffuse concerns on bias or type-1 error, may also result in false decisions (in either direction) \citep{Berger1985DecisionTheory}. Notably, those false decisions are introduced through lack of information, resulting in wide confidence intervals, with resulting claims that "no decision can be made". Understandingly, not making a decision due to "too little information" results in passive decision making. An active decision could still have been made, yet at greater risk.

Assumptions are ubiquitous in pharmaceutical statistics and commonly used.
Understanding of uncertainties and assumptions may seem less relevant in the analysis of sufficiently powered confirmatory clinical trials, which report p-values based on non-parametric or semi-parametric statistical analyses (e.g. Cox proportional hazards (PH) model \citep{Cox1972}, proportional odds assumption in ordinal logistic regression \citep{McCullagh1980Ordinal},  MMRM \citep{Mallinckrodt2008} and stratified analyses in general \citep{Senn2024Stratification}). However, even in those situations, modeling and assumptions may impact decision-making, for example, through the strategy of handling intercurrent events, which is typically defined in the Estimand, the estimation of treatment effects in adaptive designs \citep{Robertson2023AdaptiveEstimation} or more generally the estimation of treatment and covariate effects in non-linear models \citep{Wei2023ConditionalMarginal}.

Quantification of uncertainty is a key responsibility for statisticians. While non-parametric statistical approaches may reduce risk of bias, they may also disregard important information, which could support decision-making. The validity of the extra information depends on the appropriateness of the model. As \citet{box1979robustness} stated: all models are wrong - but some are useful. Complex models and innovative design approaches, which integrate assumptions and external data sources may fail to reach acceptance due to non-validated concerns about bias and type-1 error control, while their application may be fully appropriate in {\it specific contexts} to inform decisions at risk. 

Within this paper, we discuss the use of the ICH M15 framework \citep{ICHM15_2026} for Model-Informed-Drug-Development (MIDD) as a tool to support discussion and assessment of assumptions and modeling for informing decision making in Pharmaceutical Drug Development. We argue that the framework provides helpful guidance to justify, assess and validate the use of innovative modeling and statistical methodology for case specific intended contexts of use. If properly implemented by statisticians, ICH M15 may increase transparency on assumptions and risks in drug development, and may effectively guide discussions across stakeholders.

\section{Assumptions, modeling and decisions: Supporting frameworks}
One of the aim of statistics is to inform decisions based on data. Decisions could be made by relying on data from well designed experiments only, they could be made by relying on a blend of historical data, designed experiments and clinical experience, and some decisions need to be made without availability of data at all. The approach to decision-making and the statistics used for making a decision depends naturally on the underlying scientific problem, the opportunities and risks implied by a decision and the specific decision context. 

From this perspective, it is important to carefully select the statistical tools to inform decisions based on their suitability for the case specific context of use. Modeling assumptions are frequently made to inform decisions efficiently. Examples for assumptions include the proportional hazards assumption in survival modeling \citep{Cox1972} and the assumption of constant treatment effect for use of non-concurrent control data in platform trials \citep{BofillRoig2026NCC}. While it is easy to challenge the validity of such assumptions, arguing for introduction of bias and inflation of type-1 error, those assumptions may still be useful in specific contexts. Beyond expressing concerns on potential invalidities, evaluation of the quantitative risk of invalidities, their influence on decisions and the ultimate implication to the affected stakeholders need to be assessed. Multiple regulatory guidance documents over the past decade introduced contextually similar frameworks for assessing appropriateness of using assumptions and quantitative approaches for answering research questions  \citep{fda2021device, fda2023ai, ich2024m15}. Those guidances have at least partly been informed by common considerations, published in a standard of the American Society of Mechanical Engineers \citep{asme2018v&v40} on assessing credibility of computational modeling through verification and validation in medical devices. Eventually, this standard was highly referenced in the FDA guidance on modeling and simulation in device submissions \citep{fda2021device} and in the MIDD credibility assessment framework introduced by \citet{Marshall2019MIDD,kuemmel2020considerations}, which is a precursor to the 2026 ICH M15 guidance on MIDD \citep{ICHM15_2026}. The frameworks have in common that they start with the formulation of a clear scientific question and a specific context of use for the proposed modeling approach. The frameworks include a transparent description of modeling influence and risks. Deviations from common regulatory practice should be justified and implications be made transparent. A key component of all frameworks includes well defined validation and verification of the models, including the documentation of those activities to increase transparency and reproducibility. 

In a slightly different manner, \citet{Benda2010ScenarioEvaluation} introduced a framework for scenario planning to support clinical development strategies, including in particular innovative design elements (seamless Phase 2/3), on a program, trial or analysis level. Similar to the earlier named frameworks, \citet{Benda2010ScenarioEvaluation} focus the evaluation around a key objective or research question. Different design/method alternatives are proposed and evaluated on a range of clinical assumptions, and a decision is made on the preferred option based on a quantitative summary, such as scientific validity (e.g. type-1 error), sensitivity, robustness or probability of success. However, the framework by \citet{Benda2010ScenarioEvaluation} is less specific on aspects such as the context of use, risks and verification/validation as compared to the above mentioned MIDD frameworks developed subsequently.

A parallel development for increasing transparency in communication of assumptions and their implications in statistical modeling took place in the last decade with the introduction of the Estimand framework \citep{ich2019e9} and a growing uptake of causal inference approaches for properly addressing intercurrent events. In contrast to the earlier mentioned modeling frameworks, the Estimand framework aims for transparency in the clinical question of interest for a specific clinical trial. Different approaches of handling of intercurrent events (e.g. change of medication), will typically result in different estimators of the treatment effect and more importantly, they result in different interpretations of the finally reported treatment effect. Introduction of the Estimand framework aimed to increase transparency in the scientific question of interest, intercurrent event handling and the resulting estimator. Interpretation of the analysis should be clarified, by using a well-defined Estimand. Ultimately, the scientific question should guide study design and the intercurrent event handling, while the intercurrent event handling would subsequently define the estimator \citep{Polverejan2023Estimands}. Similar to the situations of the modeling frameworks, the choice of estimator for addressing intercurrent events may depend on working assumptions, which may invalidate the Estimand \citep{vanlancker2025estimands}. Such issues could sometimes be addressed by changing the question of interest. However, as the question of interest has typically been selected on purpose, accepting the potential concern and providing analyses properly assessing the validity of the concern and the implication to resulting decisions should equally be considered. 

Closely related to the Estimand framework is the PICOS framework, which is used in health technology assessment (HTA) and reimbursement decision-making to define evidence requirements in terms of population, intervention, comparators, outcomes and study design \citep{Richardson1995PICO,CochranePICO2024}. While helpful for designing clinical trials, the PICOS framework is out of scope for this paper, as it does not consider modeling assumptions, but can be rather summarized as a framework providing clear design requirements by HTAs to drug developers. 

Growing application of causal inference methods in pharmaceutical drug development is also driven over the last years through the increased use of real world data and the aim of deriving real world evidence to support trial design, analyses and HTA submissions. Causal inference approaches aim for valid and efficient inference under the condition of no-unmeasured confounding. This assumption of no unmeasured confounding is an untestable assumption \citep{hernan2020causal} and sensitivity analyses, such as the E-Value \citep{vandertweel2021e-value}(effect an unmeasured confounder would need to have to explain the observed treatment effect) are commonly used to increase confidence in the research findings. Transparency and documentation are of key importance to assert credibility of analyses based on real world data \citep{FDA2023RWE,EMA2024RWE} and to reduce the risk that erroneous conclusions and contradicting results for the same research questions are being published \citep{Hernan2008Hormone}. The introduction of frameworks such as the Target Trial Framework \citep{hernan2016target}, the Causal Roadmap \citep{Nance2024CausalRoadmap}, Causal Questions and Principled Answers \citep{Goetghebeur2020CausalQuestions}, or Four Targets \citep{Lu2025FourTargets} all try to formalize the processes to reliable analyses. 

The frameworks differ from the earlier named modeling frameworks in that those frameworks are focused around a scientific research question and providing data on addressing the question in a most accurate and unbiased way. Resulting decisions based on the analyses and implications to people eventually affected by the decision are not within scope.

Bayesian modeling provides a typically assumption rich and decision centric statistical approach \citep{Spiegelhalter2004Bayesian,Ruberg2023Bayesian}. Prior distributions are used in the design and analysis stage of trials to model or quantify the uncertainty on parameters, which are relevant to a decision problem. The implications of decisions can then be quantified through loss or utility-functions which may enable immediate interpretation \citep{OHaganStevens2001CEA}. However, also those interpretations depend on underlying assumptions, including intercurrent event handling, relative weighting of responses and many others, which easily decrease transparency when interpreting the utility functions. In addition, the choice of prior distributions, has large implications on the analyses and decisions. Seemingly non-informative prior-distributions, such as a Beta(1,1) prior may eventually turn out in small sample size situations as being very informative \citep{Kerman2011Beta}. The potential value of Bayesian modeling is evident, yet it is frequently challenged due to subjectivity and transparency concerns, as the underlying statistical machinery and dependencies introduced through prior distributions may not be as easily understandable as for a non-parametric frequentist analysis \citep{Efron1986WhyNotBayesian,Gelman2008Objections}. A number of frameworks has been developed to support Bayesian applications. Numerous approaches to elicitation of prior distributions have been published in recent years \citep{oHagan2006} with applications to drug development \citep{dallow2018,Brownstein2019,Hampson2021PoS,holzhauer2022}. Those frameworks are helpful in consolidating expert input towards development of a joint understanding of scientific problems. Even without consensus on a prior distribution, the process of elicitation has large value in increasing the clarity on knowledge gaps – and sometimes in bridging those. The Bayesian workflow \citep{gelman2020bayesian} aims to support Bayesian model development. This extensive framework towards Bayesian modeling does however not address considerations of risks, implications and validity, such that concerns regarding added risks may remain after following this framework. 

Other frameworks have been published, which shall support the qualification of novel methodology \citep{EMA2017Qualification,Morris2019SimulationEvaluation}, support design and assessment of simulations \citep{Burton2006SimulationDesign,White2023CheckSimulation,Lee2026OCTAVE}, help designing sensitivity analyses \citep{Mallinckrodt2012SensitivityFramework,Norskov2021AssessmentAssumptions} or support argumentation of extrapolation \citep{EMA2018ExtrapolationReflection,ICH2024E11A}. Table \ref{tab:modelComparison} summarizes a non-exhaustive overview of frameworks, supporting assumptions and modeling in drug development. There are clearly differing objectives of frameworks. Frameworks, such as the Estimand and Target Trial framework aim to increase clarity in interpretation of analyses and to support design and analysis strategy of trials. The ICH M15, FDA AI, and Medical Device guidance integrate decision consequences into their frameworks and thereby shift focus from statistical validity to context-specific fitness of modeling approaches. This shift of focus may meaningfully increase transparency in discussions, as statistical properties, such as bias, type-1 and type-2 error would be assessed from the perspective of decision impact, severity and likelihood. 

In the following we will discuss how the ICH M15 framework may be applied to support discussion, appraisal and adoption of innovative statistical design and analysis approaches. 

\newpage
\begin{sidewaystable}[p]
\centering
\footnotesize
\caption{Frameworks for Assessing Credibility of modeling and Assumptions}
\label{tab:modelComparison}

\renewcommand{\arraystretch}{1.3}
\setlength{\tabcolsep}{6pt}

\renewcommand{\arraystretch}{1.25}
\begin{tabularx}{\linewidth}{ >{\raggedright\arraybackslash}p{4cm} X X X }
\toprule
\textbf{Framework name} & \textbf{Framework goal} & \textbf{Strength} & \textbf{Limitations} \\
\midrule
ICH M15 (MIDD)\citep{ICHM15_2026} &
Harmonize planning, evaluation, and documentation of MIDD evidence &
Focus on impact of modeling on decision and related consequences. Assessment table to follow structured evaluation and support cross-functional discussions. &
Limited practical guidance on depth of required technical information at submission stage. Focused on MIDD. \\

Bayesian Workflow \citep{gelman2020bayesian}&
Guide iterative model building, checking, and updating &
Clear process on how to fit and optimize Bayesian models, specifically related to computational issues and model comparison. &
Full focus on data analysis, with design decision and validity of analysis approach not being discussed. \\

Causal Roadmap \citep{Nance2024CausalRoadmap} &
Link causal question, data, estimation, and interpretation coherently &
Comprehensive, yet still covering full analytical process. Promotes explicit articulation of assumptions and transparency how they connect to Estimand and estimator. Encourages structured thinking. &
Recommendation to conduct simulations after data availability may be less compatible with requirements for pre-specified analysis. Technical complexity may require additional training for non-causal experts. \\

Prior Elicitation Workflow \citep{oHagan2006} &
Embed structured expert elicitation into decision making &
Increased transparency on assumptions and uncertainty in those. Process facilitates deep scientific discussions enhancing insights. &
No formal modeling framework, but a framework which could support justification of assumptions. \\

Pediatric Extrapolation (ICH E11A) \citep{ICH2024E11A} &
Harmonize framework for use of extrapolation to guide decision making in pediatrics. &
Well structured (concept, development and execution), supporting systematic evaluation and multi-disciplinary integration. &
Limited statistical depth and unclear linkage of endpoints, i.e. biomarkers vs. clinical endpoints. \\

\end{tabularx}
\end{sidewaystable}

\begin{sidewaystable}[p]
\centering
\footnotesize
\caption{Frameworks for Assessing Credibility of modeling and Assumptions}
\label{tab:frameworks_overview}

\renewcommand{\arraystretch}{1.3}
\setlength{\tabcolsep}{6pt}

\renewcommand{\arraystretch}{1.25}
\begin{tabularx}{\linewidth}{ >{\raggedright\arraybackslash}p{4cm} X X X }
\toprule
\textbf{Framework name} & \textbf{Framework goal} & \textbf{Strength} & \textbf{Limitations} \\
\midrule
Simulation Study Framework \citep{Burton2006SimulationDesign} &
Ensure rigor, reproducibility, and credibility of simulations &
Structured checklist covering essential aspects of simulation designs, from objectives to evaluating performance. &
Focuses on general principles and does not delve deeply into specific complexities (e.g. Bayesian or missing data in simulations). \\

Causal Questions and Principled Answers \citep{Goetghebeur2020CausalQuestions} &
Provide compact and principled operational basis for applied causal inference &
Clear structured explanation on how to move from scientific question to causal Estimand and appropriate statistical method. &
Only focused on point exposures in prospective studies. \\

Four Targets Framework \citep{Lu2025FourTargets} &
Producing credible causal inference informing policy by specifying Estimand, population, target trial, and target validity &
Aligns with ICH E9 addendum and target trial framework. Validity and population target increase transparency. &
Target validity setup lacks explicit or operational framework for assessing validity in practice.\\

Target Trial Emulation\\ \citep{hernan2016target} &
Emulate randomized trials using observational data &
Clarifies causal contrasts and design assumptions &
Requires strong assumptions and rich data \\

Generalizability \& Transportability \citep{Degtiar2023Generalizability}&
Used to check whether study results apply to other populations &
Formal structure for external validity, complementing internal validity checks. Encourages explicit assumptions, reducing hidden bias. Aligns with regulatory emphasis on Estimand clarity and real-world applicability &
Rich covariate data required for adjustments. Assumptions often unverifiable and may fail in practice. Computational complexity. \\

\end{tabularx}
\end{sidewaystable}

\begin{sidewaystable}[p]
\centering
\footnotesize
\caption{Frameworks for Assessing Credibility of modeling and Assumptions}
\label{tab:frameworks_overview}

\renewcommand{\arraystretch}{1.3}
\setlength{\tabcolsep}{6pt}

\renewcommand{\arraystretch}{1.25}
\begin{tabularx}{\linewidth}{ >{\raggedright\arraybackslash}p{4cm} X X X }
\toprule
\textbf{Framework name} & \textbf{Framework goal} & \textbf{Strength} & \textbf{Limitations} \\
\midrule
Estimands (ICH E9(R1)) \citep{ich2019e9} &
Framework to clarify trial objectives, design, conduct and analysis with special emphasis on intercurrent events handling and interpretation of estimators. &
Regulatory standard; Enables cross-functional design discussions. Increases transparency on interpretation of analysis. Explicit handling of intercurrent events. &
Non-standard estimators may need to be implemented for targeted Estimand. \\

PICOS \citep{Richardson1995PICO} &
Framework to define structured research questions. &
HTA standard: simple, widely used tool emphasizing early compound-agnostic design thinking. &
Conceptual tool only, without clear analysis perspective. \\
 
FDA AI Guidance (FDA (2023a))&
Framework to support AI‑derived information and data to support regulatory decisions. &
Explicit description of regulatory engagement pathways and outlines clear process; Helpful examples available in guidance &
Large overlap with ICH M15 \\

CM\&S Credibility (ASME / FDA Devices) (FDA (2021))&
Framework to improve consistency and transparency of review and increase confidence in regulatory decisions. &
Extensive list of definitions and clear process outline with extensive information for each step. Clearly spelled out need for definition of credibility factors. &
Device guidance focused to physics based mechanistic and computational models. All recommendations specific to medical devices. \\

Simulation Framework (ADEMP) \citep{Morris2019SimulationEvaluation} &
Structured guidance for designing, executing, analyzing, and reporting simulation studies in statistics. &
Clear, reproducible framework (Aims, Data, Estimands, Methods, Performance). Extension to methodology assessment straightforward. &
Limited Bayesian and no coverage of complex data-generation mechanism \\

\end{tabularx}
\end{sidewaystable}

\begin{sidewaystable}[p]
\centering
\footnotesize
\caption{Frameworks for Assessing Credibility of modeling and Assumptions}
\label{tab:frameworks_overview}

\renewcommand{\arraystretch}{1.3}
\setlength{\tabcolsep}{6pt}

\renewcommand{\arraystretch}{1.25}
\begin{tabularx}{\linewidth}{ >{\raggedright\arraybackslash}p{4cm} X X X }
\toprule
\textbf{Framework name} & \textbf{Framework goal} & \textbf{Strength} & \textbf{Limitations} \\
\midrule
OCTAVE Simulation Framework \citep{Lee2026OCTAVE}&
Guidance on how to design and conduct simulation studies. &
Integrates objectives, context of use, assumptions, validation, and decision impact &
Targeted explicitly to simulation studies. \\


Sensitivity Analyses for Missing Data \citep{Mallinckrodt2012SensitivityFramework} &
Tests robustness of conclusions under different missing data assumptions (MAR vs MNAR). &
Clear taxonomy; regulatory‑aligned; Estimand‑consistent &
Focused on missing data only \\

Statistical Analysis Methods Assessment \citep{Norskov2021AssessmentAssumptions}&
Evaluates assumptions behind regression models (e.g., linearity, normality). &
Improves model valdity and supports transparency &
Strong focus on correct modeling when data available. Limited design thinking and no discussion of implications to decisions. \\

Scenario Evaluation Framework \citep{Benda2010ScenarioEvaluation}&
Framework for planning of clinical development and analysis strategy in a structured manner, based on evaluation of scenarios. &
Clearly structured approach to compare different design or analysis options on selected scenarios. &
No emphasis on checking appropriateness of and sensitivity to assumptions. Rather a tool to pick one of multiple options available. \\
\bottomrule
\end{tabularx}
\end{sidewaystable}

\section{ICH M15 framework}
The ICH M15 guideline \citep{ICHM15_2026} provides a harmonized framework for developing, evaluating, and communicating MIDD evidence \citep{Marshall2019MIDD}. Its original scope is to establish general principles, key assessment elements, and documentation standards applicable across regulatory regions. The guideline originated from the need to unify expectations for MIDD, as prior ICH guidelines addressed only topic‑specific modeling considerations and lacked overarching principles defining acceptable use of models in regulatory decision‑making. 

Public consultation following the Step 2 draft of the guideline highlighted the need for greater clarity in distinguishing model risk from model impact, the need for more explicit guidance for artificial intelligence and machine-learning models, and the importance of aligning ICH M15 with emerging regional frameworks for advanced computational modeling. These comments reinforced a demand for a transparent, risk-based structure governing how models contribute to drug-development decisions and how their credibility should be evaluated. The final guideline reflects a light revision of the Step 2 draft \citep{ich2024m15}. The areas of model influence and consequence of a wrong decision received clarifying updates. Broader stakeholder requests, such as AI/ML guidance, expanded risk framework, practical examples, regulatory‑engagement detail, were not incorporated into the final document. 

ICH M15 discusses two different stages of the MIDD process, namely (1) the planning and regulatory interaction and (2) the implementation, reporting and evidence submission stage. Already during the planning stage, regulatory interactions are expected, to align early on acceptability of and requirements for the proposed modeling approach. While the actual analyses might not have been conducted yet (e.g. as data is still outstanding), the discussions may allow to reach alignment on a conceptual level. Once data is available and analyses are conducted, the eventual regulatory decision would be taken following the evidence submission stage. The guidance does not integrate the formal regulatory decision making, but is a guidance for applicants. As such, outcomes of models and outcomes of the MIDD assessments in the guidance do not refer to the outcome of the regulatory assessment, but to the outcome of the modeling in addressing the scientific question of interest. 

\subsection{ICH M15 as Structured framework to support statistical modeling}
To support the submission of MIDD proposals, the ICH M15 guidance provides a table for assessment of MIDD evidence, as well as the outline for a typical model analysis report. ICH M15 defines a stepwise structure that links model development and evaluation directly to regulatory decision needs. The assessment table (appendix 1 to ICH M15) iterates through the different subsections of the ICH M15 guidance, providing concise instructions for completion of the MIDD problem definition and assessment process. This introduces a structured approach to solution development, focusing on relevant discussions and comparisons.

The three sections of the ICH M15 assessment table include the key assessment elements, additional considerations for interaction with regulators to inform decision making, and considerations for the MIDD evidence submission stage. The key assessment elements summarize high level considerations on the scientific problem and proposed solution and are given as follows.
\begin{itemize}
\item[1.] \textbf{Question of Interest:}
specifies the decision the MIDD activity is intended to inform. It must be explicit and aligned with multidisciplinary and regulatory assessment needs. 
\item[2.] \textbf{Context of use:}
describes the model’s intended role, scope, underlying data sources, and all additional evidence used to address the question of interest. It defines how the model contributes to answering the regulatory question and what supporting information is required. 
\item[3.] \textbf{Model influence:} addresses the intended contribution of model outcomes within the totality of evidence, rated as low, medium or high and including a justification for the weighting.
\item[4.] \textbf{Consequence of wrong decision}: focuses on the potential harm to patient safety or efficacy resulting from an incorrect decision based on all available information (not just the proposed model), weighted as low, medium or high.
\item[5.] \textbf{Model risk:} combines the model influence and consequences of wrong decisions into a metric assessing the contribution of the model in unfavorable decision making and consequences, rated as low, medium or high.
\item[6.] \textbf{Model impact:} describes the extent by which the proposed modeling strategy deviates from regulatory standards, rated as low, medium or high.
\end{itemize}

Additional considerations for interaction with regulators outline technical criteria on the model, as well as a discussion on the appropriateness of the proposed model. The technical criteria shall inform acceptance of the modeling approach and are meant to include in particular verification and validation activities. Verification refers in ICH M15 to appropriateness of computation codes for data processing and analyses, while validation refers to the assessment of model performance and robustness. 
The assessment table closes with a summary of the MIDD evidence submission stage, including in particular the model evaluation and outcomes, as well as a discussion of the outcome of the MIDD evidence assessment. A difference between the two sections is given in the timing of completion. Technical criteria and discussion of appropriateness of the proposed model should be completed in the planning stage and optimally be aligned on with the regulators, while the last two sections summarize the eventual evidence after having completed and reviewed the planned analyses. 

ICH M15 is of high relevance for statistical modeling. It formalizes a structured decision-analytic foundation for modeling work. The framework explicitly links the question of interest, the role of the model within the totality of evidence, the consequences of incorrect decisions, and the evaluation activities needed to establish credibility. This codification aligns naturally with statistical reasoning, where model uncertainty, assumptions, data quality, and inferential consequences should be articulated and justified. The linkage between model influence, decision consequence, and required model evaluation creates a natural path for statistical justification of modeling approaches. It may enhance evidence-based decision-making in drug development, as it may increase transparency in discussion of risks, moving beyond type-1 error to include likelihood and consequences of wrong decisions into holistic decision making.
For statisticians, its emphasis on clarity of objectives, structured risk-based justification, and systematic evaluation may enhance rigorous model-based inference and effective communication of modeling results within multidisciplinary teams.

\section{Applying ICH M15 to Bayesian Borrowing}
In this section, we follow the structure of the ICH M15 assessment table and complete it for an exemplary case of dynamic Bayesian Borrowing of control data across two contemporaneous trials in Multiple Myeloma. The objective of the section is not to document the full technical implementation, which would rather be covered in the modeling report, but to show how the ICH M15 assessment table helps clarify what decision is being informed, how much the model matters, what could go wrong, and what level of evidence is proportionate to the associated risks.

\subsection{Key Assessment Elements}
Here, we present the key assessment elements as executive-level summary of the modeling proposal, as suggested in the ICH M15 assessment table. The investigational treatments (TRT) are here assumed to be called TRT-A and TRT-B, while the control group of interest is abbreviated as CTRL-A.
\subsubsection{Question of Interest}
The question of interest defines the decision to be informed, which is independent of modeling choices or data sources:
\begin{center}
\begin{minipage}{0.85\textwidth}
Does regimen TRT-A improve overall survival compared with CTRL-A in patients with relapsed or refractory multiple myeloma who received 1–3 prior lines of therapy?
\end{minipage}
\end{center}
This framing of the questions makes explicit that the modeling – here the Bayesian borrowing – is not the actual objective, but just one approach for informing the decision.

\subsubsection{Context of Use}
The context of use specifies the explicit situation and how modeling contributes to answering the question, including its scope and supporting evidence.

\begin{center}
\begin{minipage}{0.85\textwidth}
Two confirmatory trials in overlapping patient populations and with a common control regimen are planned. Trial A evaluates TRT-A versus CTRL-A, while Trial B evaluates TRT-B versus CTRL-A. Each trial is independently powered for the primary endpoint of progression free survival (PFS). Overall survival (OS) is a key secondary endpoint of high relevance to regulators, patients, and payers, but is expected to mature substantially later.
Using the proposed approach, CTRL-A data on OS from Trial B would be incorporated into the OS analysis of Trial A, by implementing Bayesian dynamic borrowing, with propensity score weighting to address residual population differences at baseline. This approach is intended to:
\begin{itemize}
\item	inform OS decisions earlier,
\item	reduce uncertainty in treatment effect estimation on OS, and
\item	improve the ability to detect both benefit and potential detriment.
\end{itemize}
\end{minipage}
\end{center}
Using ICH M15, it is encouraged to explicitly communicate what the model does and what it does not do. In the given case, the dynamic borrowing augments and does not replace trial internal evidence. It is conditional on established PFS benefit.

\subsubsection{Model Influence}
Model influence reflects the relative weight of the model within the totality of evidence and is here not defined by quantitative statistical properties.
\begin{center}
\begin{minipage}{0.85\textwidth}
Model influence is assessed as medium:
\begin{itemize}
\item	Trial internal randomized data remain the primary evidence source.
\item	The model informs decision making only after success on the primary endpoint.
\item	External data are dynamically down weighted when data-conflicts are detected.
\end{itemize}
Sensitivity analyses, including standalone analyses and internal–external control comparisons across multiple endpoints, further limit the circumstances under which the model materially would alter conclusions.
\end{minipage}
\end{center}
This explicit articulation of influence helps non-statistical stakeholders understand when the model matters, and when it does not.

\subsubsection{Consequence of a Wrong Decision}
Consequence refers to the severity of patient level and decision level harm if an incorrect conclusion is drawn, considering all information.
\begin{center}
\begin{minipage}{0.85\textwidth}
For OS, consequences of wrong decisions are assessed as high, irrespective of the modeling strategy. Potential errors include:
\begin{itemize}
\item	falsely claiming survival benefit,
\item	failing to identify true benefit,
\item	missing or falsely signaling harm,
\item	and delaying correct decisions due to immature data.
\end{itemize}
\end{minipage}
\end{center}

With the identification of “consequences of wrong decisions”, and the term “all information”, ICH M15 intentionally separates severity of consequences from the methodological risk. This helps guiding an open discussion on model properties and preventing the false assumption that “conservative” methods are always less risky.

\subsubsection{Model Risk}
Model risk finally combines model influence and decision consequence into an overall qualitative assessment.
\begin{center}
\begin{minipage}{0.85\textwidth}
Given medium model influence and high decision consequence, the model risk is considered medium. The proposed approach may increase the probability of early conclusions, but also reduces the risk of delayed detection of harm or benefit. The risks are evaluated relative to the alternative of relying on immature standalone data, rather than in isolation.
\end{minipage}
\end{center}
This framing shifts discussion from abstract concerns (e.g. bias and type-1 error) towards comparative decision risk.

\subsubsection{Model Impact}
Model impact describes the degree, with which the proposed strategy deviates from conventional regulatory practice.
\begin{center}
\begin{minipage}{0.85\textwidth}
Bayesian dynamic borrowing represents a methodological extension beyond standard frequentist analyses with the aim of supporting decisions by external data. The incremental impact is assessed as low because:
\begin{itemize}
\item   the Estimand remains unchanged,
\item	external data augment, but do not dominate inference,
\item	multiple safeguards are implemented to strengthen exchangeability and robustness,
\item	and key operating characteristics are prespecified and evaluated,
\end{itemize}
\end{minipage}
\end{center}
Distinguishing novelty from incremental risk through ICH M15 provides an important opportunity to clarify cross functional discussions.

\subsection{ICH M15 Planning Stage: Structuring Early Alignment}
Bayesian borrowing requires substantial upfront evaluation to calibrate models and understand operating characteristics. ICH M15 provides a way to position this work during the planning stage, focusing discussion on acceptability criteria rather than technical detail.
\subsubsection{Technical Criteria (Decision Relevant Conditions)}
Instead of viewing technical criteria as post hoc validation, they can be framed as conditions under which model based evidence is considered admissible for decision making and can thereby actively inform study design. 

\begin{center}
\begin{minipage}{0.85\textwidth}
The following technical criteria shall be implemented to mitigate concerns on borrowing between internal and external OS control data: 
\begin{itemize}
\item   absence of severe discrepancy between internal and external control data on key endpoints (ORR, PFS and OS in particular), assessed using two-sided tests at level $20\%$,
\item	limiting the effective amount of borrowed information (total effective number of OS events should not exceed $150\%$ of Trial A events),
\item	evidence of a favorable trend in standalone data, demonstrated by exceeding $80\%$ conditional power for a stand-alone final OS analysis,
\item	and acceptable control of type-1 error under relevant scenarios, achieved via stand-alone p-value for OS below $10\%$ one-sided.
\end{itemize}
\end{minipage}
\end{center}
The technical criteria allow clinical, regulatory, and statistical stakeholders to align early on what would invalidate the approach, thereby supporting transparent governance. Justification, calibration and specification of technical criteria would optimally be provided in the model analysis report.

\subsubsection{Appropriateness of the Proposed MIDD}
An explicit justification of why the modeling approach is appropriate for its context is required by ICH M15. 

\begin{center}
\begin{minipage}{0.85\textwidth}
The application of Bayesian Dynamic Borrowing for assessing OS benefits is considered appropriate due to the following reasons:
\begin{itemize}
\item	the long maturation time of OS relative to decision needs,
\item	concurrent trials with overlapping sites, populations and control regimens,
\item	pre-specified methods to address residual differences in populations,
\item	similarity of internal and external control group data on three endpoints (OS, PFS and ORR),
\item	and decision conditional on having established superiority on the primary endpoint.
\end{itemize}
\end{minipage}
\end{center}
This discussion focuses on decision logic and does not focus on computational detail. 

\subsection{Evidence Submission Perspective: Communicating Outcomes Succinctly}
While the evidence submission stage formally targets regulatory submissions, the proposed structure and considerations are equally relevant for informing internal communication.
\subsubsection{Model Evaluation (Summary View)}
At the assessment table level, the model evaluation is summarized through a qualitative synthesis of operating characteristics across relevant scenarios, including identification of risks, and comparison with alternatives. Details on the operating characteristics are stored and documented in the modeling report.
\subsubsection{Outcome of Model Evaluation}
The outcome statement summarizes in a concise manner the multidisciplinary judgment on the proposed modeling approach, i.e. whether it is sufficiently credible and proportionate for its intended use.
In the illustrated case, dynamic borrowing modestly improves the probability of timely OS conclusions, while additional risks remain limited and controlled relative to both standalone and naïve pooling approaches.

\subsection{Role of ICH M15 in Cross Functional modeling Discussions}
The discussed example illustrates that a primary value of ICH M15 for statistical modeling lies not in prescribing methods, but in structuring dialogue and focusing communication. By anchoring discussion around:
\begin{itemize}
\item	decision purpose,
\item	influence of modeling,
\item	severity of consequences,
\item	proportionality of evidence,
\end{itemize}
the framework helps multidisciplinary teams move beyond binary questions of “acceptable” versus “unacceptable” methods toward a shared understanding of decision risk. This shift is essential for Bayesian Dynamic Borrowing and other assumption rich modeling strategies, to enable transparent, defensible decision making.
It is understood that the ICH M15 process will in most cases be iterative. For example, stakeholders may identify additional consequences of wrong decisions (e.g. multiple false positives), or findings in the evaluation may result in study design or acceptability criteria adjustments (e.g. 2:1 randomization instead of 1:1). Those iterations do not limit, but actually demonstrate the value of ICH M15 in prospectively and transparently addressing all stakeholders’ concerns when answering the question of interest.

\section{Discussion}
Statistical modeling plays a central role in pharmaceutical drug development by shaping decisions under uncertainty. Many development and regulatory decisions rely - explicitly or implicitly - on modeling assumptions, external data, and synthesis of heterogeneous evidence. These assumptions are often treated asymmetrically: conservative, assumption lean approaches are readily accepted, sometimes without full understanding of still existing underlying assumptions and limitations. More explicit or innovative modeling strategies face heightened scrutiny due to concerns about bias, type I error inflation, or interpretability. As argued throughout this paper, such asymmetry risks obscuring an equally important source of error: the consequences of not using relevant information when it is available and scientifically justified.

Against this background, the ICH M15 guideline represents an important conceptual advance. By explicitly linking the question of interest, context of use, model influence, and consequences of incorrect decisions, the framework shifts the evaluation of modeling approaches from a narrow focus on statistical properties toward an assessment of decision relevance and fitness for purpose. Although originally developed for mechanistic models, this structure is directly applicable to statistical modeling strategies commonly encountered in drug development, including Bayesian borrowing, integration of external data, and parametric longitudinal analyses.

A central contribution of ICH M15 is its explicit framing of model risk as a function of both evidentiary influence and decision consequence. Traditional statistical metrics such as bias or type I error remain essential, but they are insufficient in isolation to guide decisions in contexts where false negatives, delayed conclusions, or failure to detect harm carry substantial implications for patients. In such settings, dismissing a modeling approach solely because it relies on assumptions may increase overall decision risk rather than reduce it. This perspective is particularly relevant given that assumptions are pervasive in statistical practice, including many that are non testable and routinely accepted without formal assessment of their implications. ICH M15 requires that such assumptions be articulated and evaluated relative to their intended context of use, with validation and sensitivity analyses calibrated to their potential impact on decisions. Importantly, the framework emphasizes proportionality: higher risk decisions warrant more extensive evaluation, whereas lower risk applications may justify pragmatic approaches. As noted by \citet{Burman2005DecisionMaking}, excessive rigor in pursuit of theoretical optimality may itself be harmful if it prevents timely or informative decisions. The relevant question is therefore not whether a model is perfect, but whether it is sufficiently credible to inform a specific decision, given its influence and the consequences of error.

Finally, the framework provides a common language for cross functional and regulatory dialogue. By focusing discussions on model influence, decision consequences, and credibility requirements, it shifts attention away from abstract methodological debates toward concrete, decision relevant questions: What decision is being informed? What could go wrong? How severe would the consequences be? And what evidence is required to establish sufficient confidence? This structured approach is particularly valuable for innovative designs and analyses that fall outside established regulatory precedent.

Further experience is needed to clarify how decision centric considerations - such as timing, false negative risk, and patient impact - can be systematically balanced with traditional operating characteristics within a unified assessment. Consolidating lessons learned from real regulatory applications will be essential to promote consistency, shared expectations, and effective cross functional dialogue.

In summary, the ICH M15 framework offers statisticians a principled and defensible approach to justify modeling strategies in pharmaceutical drug development. Applied thoughtfully, it supports transparent communication of assumptions and risks, encourages proportional evaluation of model credibility, and facilitates informed, decision focused use of statistical models in regulatory settings.

\section*{Acknowledgments}
This work was initiated and supported by the EFSPI Methodology Leaders Group.

\section*{Funding}
This research received no funding.

\section*{Ethics Statements}
The authors have nothing to report.

\section*{Conflicts of Interest}
Marc Vandemeulebroecke is Editor in Chief for Pharmaceutical Statistics. There are no other conflicts of interest to report.

\section*{Data Availability Statement}
Data sharing not applicable to this article as no datasets were generated or analysed during the current study.

\bibliographystyle{chicago}

\bibliography{modellingAndRisks.bib}

@article{burger2021external,
  title={The use of external controls: To what extent can it currently be recommended?},
  author={Burger, Hans U and Gerlinger, Christoph and Harbron, Chris and Koch, Armin and Posch, Martin and Rochon, Justine and Schiel, Armin},
  journal={Pharmaceutical Statistics},
  volume={20},
  number={6},
  pages={1002--1016},
  year={2021},
  publisher={Wiley Online Library}
}

@article{kuemmel2020considerations,
  author  = {Kuemmel, C. and Yang, Y. and Zhang, X. and Florian, J. and Zhu, H.
             and Tegenge, M. and Huang, S.-M. and Wang, Y. and Morrison, T.
             and Zineh, I.},
  title   = {Consideration of a Credibility Assessment Framework in Model-Informed Drug Development:
             Potential Application to Physiologically-Based Pharmacokinetic Modeling and Simulation},
  journal = {CPT: Pharmacometrics \& Systems Pharmacology},
  year    = {2020},
  volume  = {9},
  number  = {1},
  pages   = {21--28},
  doi     = {10.1002/psp4.12479},
  pmid    = {31652029}
}

@article{Brownstein2019,  
  author  = {Brownstein, Naomi C. and Louis, Thomas A. and O'Hagan, Anthony and Pendergast, Jane},
  title   = {The Role of Expert Judgment in Statistical Inference and Evidence-Based Decision-Making},
  journal = {The American Statistician},
  year    = {2019},
  volume  = {73},
  number  = {S1},
  pages   = {56--68},
  doi     = {10.1080/00031305.2018.1529623},
  publisher = {Taylor \& Francis}
}

@book{box1979robustness,
  title={Robustness in the strategy of scientific model building},
  author={Box, George EP},
  year={1979},
  publisher={Academic Press},
  address={New York},
  doi = {https://doi.org/10.1016/B978-0-12-438150-6.50018-2}
}

@techreport{asme2018v&v40,
  title={{ASME V\&V 40-2018: Assessing Credibility of Computational Modeling through Verification and Validation: Application to Medical Devices}},
  author={{ASME}},
  institution={American Society of Mechanical Engineers},
  year={2018}
}

@misc{fda2023ai,
  title={{Artificial Intelligence/Machine Learning (AI/ML)-Based Software as a Medical Device (SaMD) Action Plan}},
  author={{U.S. Food and Drug Administration}},
  year={2023},
  howpublished={U.S. Food and Drug Administration}
}

@misc{fda2021device,
  title={{Assessing the Credibility of Computational Modeling and Simulation in Medical Device Submissions}},
  author={{U.S. Food and Drug Administration}},
  year={2021},
  howpublished={U.S. Food and Drug Administration, Draft Guidance}
}

@article{hernan2016target,
  title={Using big data to emulate a target trial when a randomized trial is not available},
  author={Hern{\'a}n, Miguel A and Robins, James M},
  journal={American Journal of Epidemiology},
  volume={183},
  number={8},
  pages={758--764},
  year={2016},
  publisher={Oxford University Press}
}

@book{hernan2020causal,
  title={Causal Inference: What If},
  author={Hern{\'a}n, Miguel A and Robins, James M},
  year={2020},
  publisher={Chapman \& Hall/CRC},
  address={Boca Raton}
}

@article{vanlancker2025estimands,
  title={Chasing Shadows: How Implausible Assumptions Skew Our Understanding of Causal Estimands},
  author={Vansteelandt, Stijn and Van Lancker, Kelly},
  journal={Statistics In Biopharmaceutical Research},
  volume={17},
  number={4},
  pages={507--513},
  year={2025}  
}

@article{gelman2020bayesian,
  title={Bayesian workflow},
  author={Gelman, Andrew and Vehtari, Aki and Simpson, Daniel and Margossian, Charles C and Carpenter, Bob and Yao, Yuling and Kennedy, Lauren and Gabry, Jonah and B{\"u}rkner, Paul-Christian and Modr{\'a}k, Martin},
  journal={arXiv preprint arXiv:2011.01808},
  year={2020}
}

@article{vandertweel2021e-value,
  title={Sensitivity Analysis in Observational Research: Introducing the {E}-Value},
  author={VanderWeele, Tyler J and Ding, Peng},
  journal={Annals of Internal Medicine},
  volume={167},
  number={4},
  pages={268--274},
  year={2017},
  publisher={American College of Physicians}
}

@article{bornkamp2009mcpmod,
  title={{MCPMod: An R package} for the design and analysis of dose-finding studies},
  author={Bornkamp, Bj{\"o}rn and Pinheiro, Jos{\'e} and Bretz, Frank},
  journal={Journal of Statistical Software},
  volume={29},
  pages={1--23},
  year={2009}
}

@article{bretz2005combining,
  title={Combining multiple comparisons and modeling techniques in dose-response studies},
  author={Bretz, Frank and Pinheiro, Jos{\'e} C and Branson, Michael},
  journal={Biometrics},
  volume={61},
  number={3},
  pages={738--748},
  year={2005},
  publisher={Oxford University Press}
}

@article{dallow2018,
  title={Better decision making in drug development through adoption of formal prior elicitation},
  author={Dallow, N. and Best, N. and Montague, T.H.},
  journal={Pharmaceutical Statistics},
  volume={17},
  number={4},
  pages={301--316},
  year={2018},
  publisher={Wiley Online Library}
}

@article{holzhauer2022,
  title={Eliciting judgements about dependent quantities of interest: The SHeffield ELicitation Framework extension and copula methods illustrated using an asthma case study},
  author={Holzhauer, Björn and Hampson, Lisa V. and Gosling, John P. and Bornkamp, Björn and Kahn, Joseph and Lange, Markus R. and Luo, Wen-Lin and Brindicci, Caterina and Lawrence, David and Ballerstedt, Steffen and O'Hagan, Anthony},
  journal={Pharmaceutical Statistics},
  volume={21},
  number={5},
  pages={1005--1021},
  year={2022},
  publisher={Wiley Online Library}
}

@book{oHagan2006,
      author = "O'Hagan, Anthony and Buck, Caitlin E. and Daneshkhah, Alizera and Eiser, J. Richard and Garthwaite, Paul H. and Jenkinson, David J. and Oakley, Jeremy E. and Rakow, Tim",
       title = "Uncertain Judgements: Eliciting Experts' Probabilities",
        publisher = "John Wiley \& Sons, Ltd",
        year = "2006"
}

@article{Sertkaya24,
  author  = {Sertkaya, A. and Beleche, T. and Jessup, A. and Sommers, B.D.},
  title   = {Costs of Drug Development and Research and Development Intensity in the US, 2000-2018.},
  journal = {JAMA Network Open},
  year    = {2024},
  volume  = {7},
  number  = {6},
  doi     = {doi:10.1001/jamanetworkopen.2024.15445}
}

@book{ChuangSteinKirby2021,
  author    = {Chuang-Stein, Christy and Kirby, Simon},
  title     = {Quantitative Decisions in Drug Development},
  edition   = {1},
  year      = {2017},
  publisher = {Springer International Publishing},
  address   = {Cham, Switzerland},
  series    = {Springer Series in Pharmaceutical Statistics},
  isbn      = {978-3-030-79730-0},
  doi       = {10.1007/978-3-030-79731-7}
}

@article{OHagan2005Assurance,
  author  = {O'Hagan, Anthony and Stevens, John W. and Campbell, Michael J.},
  title   = {Assurance in clinical trial design},
  journal = {Pharmaceutical Statistics},
  year    = {2005},
  volume  = {4},
  number = {3},
  pages   = {187--201},
  doi     = {10.1002/pst.175}
}

@book{JennisonTurnbull1999,
  author    = {Jennison, Christopher and Turnbull, Bruce W.},
  title     = {Group Sequential Methods with Applications to Clinical Trials},
  year      = {1999},
  publisher = {Chapman and Hall/CRC},
  address   = {Boca Raton, FL},
  isbn      = {978-0-8493-0316-6},
  doi       = {10.1201/9780367805326}
}

@article{FDARareDiseases,
  author      = {{U.S. Food and Drug Administration}},
  title       = {Rare Diseases: Considerations for the Development of Drugs and Biological Products},
  institution = {FDA},
  year        = {2019},
  url         = {https://www.fda.gov/drugs/guidances-drugs/rare-diseases-considerations-development-drugs-and-biological-products}
}

@article{Viele2014HistoricalControls,
  author  = {Viele, Kert and Berry, Scott and Neuenschwander, Beat and Amzal, Billy and Chen, Fang and Enas, Nathan and Hobbs, Brian and Ibrahim, Joseph G. and Kinnersley, Nelson and Lindborg, Stacy and Micallef, Sandrine and Roychoudhury, Satrajit and Thompson, Laura},
  title   = {Use of historical control data for assessing treatment effects in clinical trials},
  journal = {Pharmaceutical Statistics},
  year    = {2014},
  volume  = {13},
  number  = {1},
  pages   = {41--54},
  doi     = {10.1002/pst.1589}
}

@article{Cox1972,
  author  = {Cox, David R.},
  title   = {Regression Models and Life-Tables},
  journal = {Journal of the Royal Statistical Society, Series B (Methodological)},
  year    = {1972},
  volume  = {34},
  number  = {2},
  pages   = {187--220}
}

@article{Wei2023ConditionalMarginal,
  author  = {Wei, Jiawei and Xu, Jiajun and Bornkamp, Bj{\"o}rn and others},
  title   = {Conditional and Unconditional Treatment Effects in Randomized Clinical Trials: Estimands, Estimation, and Interpretation},
  journal = {Statistics in Biopharmaceutical Research},
  year    = {2023},
  volume = {16},
  number = {3},
  pages = {371-381},
  doi     = {10.1080/19466315.2023.2292774}
}

@article{Robertson2023AdaptiveEstimation,
  author  = {Robertson, David S. and Choodari-Oskooei, Babak and Dimairo, Munya and Flight, Laura and Pallmann, Philip and Jaki, Thomas},
  title   = {Point Estimation for Adaptive Trial Designs {I}: {A} Methodological Review},
  journal = {Statistics in Medicine},
  year    = {2023},
  volume  = {42},
  number  = {2},
  pages   = {122--145},
  doi     = {10.1002/sim.9605}
}

@article{Verma2023QSP,
  author  = {Verma, Meghna and Gall, L. Louis and Biasetti, Jacopo and Di Veroli, Giovanni Y. and Pichardo-Almarza, Cesar and Gibbs, Megan A. and Kimko, Holly},
  title   = {Quantitative systems modeling approaches towards model-informed drug development: Perspective through case studies},
  journal = {Frontiers in Systems Biology},
  year    = {2023},
  volume  = {2},
  article = {1063308},
  doi     = {10.3389/fsysb.2022.1063308}
}

@article{FDA2018RWE,
  author      = {{U.S. Food and Drug Administration}},
  title       = {Framework for {FDA}'s Real-World Evidence Program},
  institution = {FDA},
  year        = {2018},
  url         = {https://www.fda.gov/media/120060/download}
}

@incollection{Izem2023RWE,
  author    = {Izem, Rima and Davi, Ruthanna and Luan, Jingyu Julia and Gamalo, Margaret},
  title     = {Overview of the Current Real-World Evidence Regulatory Landscape},
  booktitle = {Real-World Evidence in Medical Product Development},
  publisher = {Springer},
  year      = {2023},
  pages     = {17--28},
  doi       = {10.1007/978-3-031-26328-6_2}
}

@article{Ruberg2023Bayesian,
  author  = {Ruberg, Stephen J. and Beckers, Francois and Hemmings, Rob and Honig, Peter and Irony, Telba and LaVange, Lisa and Lieberman, Grazyna and Mayne, James and Moscicki, Richard},
  title   = {Application of {B}ayesian approaches in drug development: starting a virtuous cycle},
  journal = {Nature Reviews Drug Discovery},
  year    = {2023},
  volume  = {22},
  pages   = {235--250},
  doi     = {10.1038/s41573-023-00638-0}
}

@article{FDA2026Bayesian,
  author      = {{U.S. Food and Drug Administration}},
  title       = {Use of Bayesian Methodology in Clinical Trials of Drugs and Biological Products},
  institution = {FDA},
  year        = {2026},
  note        = {Draft Guidance for Industry},
  url         = {https://www.fda.gov/media/190505/download}
}

@article{Pocock1976,
  author  = {Pocock, Stuart J.},
  title   = {The Combination of Randomized and Historical Controls in Clinical Trials},
  journal = {Journal of Chronic Diseases},
  year    = {1976},
  volume  = {29},
  number = {3},
  pages   = {175--188},
  doi = {10.1016/0021-9681(76)90044-8}
}

@article{Freedman2005Causation,
author = {David A. Freedman},
title ={Statistical Models for Causation: What Inferential Leverage Do They Provide?},
journal = {Evaluation Review},
volume = {30},
number = {6},
pages = {691-713},
year = {2006},
doi = {10.1177/0193841X06293771},
    note ={PMID: 17093105},
URL = {https://doi.org/10.1177/0193841X06293771},
eprint = {https://doi.org/10.1177/0193841X06293771}
}

@book{Berger1985DecisionTheory,
  author    = {Berger, James O.},
  title     = {Statistical Decision Theory and Bayesian Analysis},
  publisher = {Springer},
  edition = {2nd},
  address   = {New York, NY},
  year      = {1985},
  doi={10.1007/978-1-4757-4286-2}
}

@article{McCullagh1980Ordinal,
  author  = {McCullagh, Peter},
  title   = {Regression Models for Ordinal Data},
  journal = {Journal of the Royal Statistical Society: Series B (Methodological)},
  year    = {1980},
  volume  = {42},
  number  = {2},
  pages   = {109--142}
}

@article{Senn2024Stratification,
  author  = {Senn, Stephen and K{\"o}nig, Franz and Posch, Martin},
  title   = {Stratification in Randomised Clinical Trials for Rare Diseases and Analysis of Covariance: Some Simple Theory and Recommendations},
  journal = {arXiv},
  year    = {2024},
  note    = {arXiv:2408.06760}
}

@article{Mallinckrodt2008,
  author       = {Mallinckrodt, CH and Lane, PW and Schnell, D and Peng, Y and Mancuso, JP},
  title        = {Recommendations for the Primary Analysis of Continuous Endpoints in Longitudinal Clinical Trials},
  journal = {Drug information journal: DIJ/Drug Information Association},
  year         = {2008},
  volume = {42},
  number = {4},
    pages = {303-319},
    doi = {10.1177/009286150804200402}
}

@article{BofillRoig2026NCC,
  author  = {Bofill Roig, Marta and Krotka, Pavla and Hees, Katharina and Koenig, Franz and Magirr, Dominic and Jacko, Peter and Parke, Tom and Posch, Martin},
  title   = {Treatment--Control Comparisons in Platform Trials Including Non-Concurrent Controls},
  journal = {Statistics in Biopharmaceutical Research},
  year    = {2026},
  volume  = {18},
  number  = {1},
  pages   = {79--96},
  doi     = {10.1080/19466315.2025.2496316}
}

@article{Polverejan2023Estimands,
  author  = {Polverejan, Elena and O'Kelly, Michael and Hefting, Nanco and Norton, Jonathan D. and Lim, Pilar and Walton, Marc K.},
  title   = {Defining Clinical Trial {E}stimands: A Practical Guide for Study Teams with Examples Based on a Psychiatric Disorder},
  journal = {Drug information journal: DIJ/Drug Information Association},
  year    = {2023},
  volume  = {57},
  numer = {5},
  pages   = {911--939},
  doi     = {10.1007/s43441-023-00524-2}
}

@article{Richardson1995PICO,
  author  = {Richardson, WS and Wilson, MC and Nishikawa, J and Hayward, RS},
  title   = {The Well-Built Clinical Question: A Key to Evidence-Based Decisions},
  journal = {ACP Journal Club},
  year    = {1995},
  volume  = {123},
  number = {3},
  pages   = {A12-A13},
  doi     = {10.7326/ACPJC-1995-123-3-A12}
}

@inbook{CochranePICO2024,
  author    = {McKenzie, Joanne E. and Brennan, Sue E. and Ryan, Rebecca E. and Thomson, Hilary J. and Johnston, Renea V. and Thomas, James},
  title     = {Defining the criteria for including studies and how they will be grouped for the synthesis},
  booktitle = {Cochrane Handbook for Systematic Reviews of Interventions},
  publisher = {John Wiley \& Sons, Ltd},
    chapter = {3},
    pages = {33-65},
    year      = {2019},
  doi = {https://doi.org/10.1002/9781119536604.ch3},
  url = {https://onlinelibrary.wiley.com/doi/abs/10.1002/9781119536604.ch3}
}

@article{FDA2023RWE,
  author      = {{U.S. Food and Drug Administration}},
  title       = {Considerations for the Use of Real-World Data and Real-World Evidence to Support Regulatory Decision-Making for Drug and Biological Products},
  institution = {FDA},
  year        = {2023},
  url         = {https://www.fda.gov/regulatory-information/search-fda-guidance-documents/considerations-use-real-world-data-and-real-world-evidence-support-regulatory-decision-making-drug}
}

@article{EMA2024RWE,
  author      = {{European Medicines Agency}},
  title       = {Real-World Evidence Provided by {EMA}: Support for Regulatory Decision-Making},
  institution = {EMA},
  year        = {2024},
  number      = {EMA/152628/2024},
  url         = {https://www.ema.europa.eu/en/documents/other/guide-real-world-evidence-provided-ema-support-regulatory-decision-making_en.pdf}
}

@article{Hernan2008Hormone,
  author  = {Hern{\'a}n, Miguel A. and Alonso, Alvaro and Logan, Roger and Grodstein, Francine and Michels, Karin B. and Willett, Walter C. and Manson, JoAnn E. and Robins, James M.},
  title   = {Observational Studies Analyzed Like Randomized Experiments: An Application to Postmenopausal Hormone Therapy and Coronary Heart Disease},
  journal = {Epidemiology},
  year    = {2008},
  volume  = {19},
  number  = {6},
  pages   = {766--779},
  doi     = {10.1097/EDE.0b013e3181875e61}
}

@article{Spiegelhalter2004Bayesian,
  author  = {Spiegelhalter, David J.},
  title   = {Incorporating Bayesian Ideas into Health-Care Evaluation},
  journal = {Statistical Science},
  year    = {2004},
  volume  = {19},
  number  = {1},
  pages   = {156--174},
  doi     = {10.1214/088342304000000080}
}

@article{Kerman2011Beta,
  author  = {Kerman, J.},
  title   = {Neutral Noninformative and Informative Conjugate Beta and Gamma Prior Distributions},
  journal = {Electronic Journal of Statistics},
  year    = {2011},
  volume  = {5},
  pages   = {1450--1470},
  doi     = {10.1214/11-EJS650}
}

@article{OHaganStevens2001CEA,
  author  = {O'Hagan, Anthony and Stevens, John W.},
  title   = {A Framework for Cost-Effectiveness Analysis from Clinical Trial Data},
  journal = {Health Economics},
  year    = {2001},
  volume  = {10},
  number = {4},
  pages   = {303--315},
    doi = {10.10002/hec.617}
}

@article{Efron1986WhyNotBayesian,
  author  = {Efron, Bradley},
  title   = {Why Isn't Everyone a {B}ayesian?},
  journal = {The American Statistician},
  year    = {1986},
  volume  = {40},
  number  = {1},
  pages   = {1--5}
}

@article{Gelman2008Objections,
  author  = {Gelman, Andrew},
  title   = {Objections to {B}ayesian Statistics},
  journal = {Bayesian Analysis},
  year    = {2008},
  volume  = {3},
  number  = {3},
  pages   = {445--450}
}

@article{Hampson2021PoS,
  author  = {Hampson, Lisa V. and Bornkamp, Bj{\"o}rn and Holzhauer, Bj{\"o}rn and Kahn, Joseph and Lange, Markus R. and Luo, Wen-Lin and Della Cioppa, Giovanni and Stott, Kelvin and Ballerstedt, Steffen},
  title   = {Improving the Assessment of the Probability of Success in Late-Stage Drug Development},
  journal = {Pharmaceutical Statistics},
  year    = {2021},
  volumen = {21},
  number    ={2},
  pages ={439--459},
  doi = {10.1002/pst.2179}
}

@article{Morris2019SimulationEvaluation,
  author  = {Morris, Tim P. and White, Ian R. and Crowther, Michael J.},
  title   = {Using Simulation Studies to Evaluate Statistical Methods},
  journal = {Statistics in Medicine},
  year    = {2019},
  volume  = {38},
  number  = {11},
  pages   = {2074--2102},
  doi     = {10.1002/sim.8086}
}

@article{Lee2026OCTAVE,
  author  = {Lee, Kim May and Choodari-Oskooei, Babak and Grayling, Michael J. and Jacko, Peter and Kimani, Peter K. and Mukherjee, Aritra and Pallmann, Philip and Parke, Tom and Robertson, David S. and Wang, Ziyan and Yap, Christina and Jaki, Thomas},
  title   = {Clinical Trial Simulation: Planning with the OCTAVE Framework, Implementation and Validation Principles},
  journal = {Statistics in Medicine},
  year    = {2026},
  volume  = {45},
  number  = {6-7},
  doi     = {10.1002/sim.70449}
}

@article{EMA2017Qualification,
  author      = {{European Medicines Agency}},
  title       = {Essential Considerations for Successful Qualification of Novel Methodologies},
  institution = {European Medicines Agency},
  year        = {2017},
  number      = {EMA/750178/2017},
  address     = {London, UK},
  url         = {https://www.ema.europa.eu/en/documents/other/essential-considerations-successful-qualification-novel-methodologies_en.pdf}
}

@article{EMA2018ExtrapolationReflection,
  author      = {{European Medicines Agency}},
  title       = {Adopted Reflection Paper on the Use of Extrapolation in the Development of Medicines for Paediatrics -- Revision 1},
  institution = {European Medicines Agency},
  year        = {2018},
  number      = {EMA/189724/2018},
  address     = {London, United Kingdom},
  url         = {https://www.ema.europa.eu/en/documents/scientific-guideline/adopted-reflection-paper-use-extrapolation-development-medicines-paediatrics-revision-1_en.pdf}
}

@article{White2023CheckSimulation,
  author  = {White, Ian R. and Pham, Tra My and Quartagno, Matteo and Morris, Tim P.},
  title   = {How to Check a Simulation Study},
  journal = {International Journal of Epidemiology},
  year    = {2024},
  volume  = {53},
  number  = {1},
  doi     = {10.1093/ije/dyad134}
}

@article{Burton2006SimulationDesign,
  author   = {Burton, Andrea and Altman, Douglas G. and Royston, Patrick and Holder, Roger L.},
  title   = {The Design of Simulation Studies in Medical Statistics},
  journal = {Statistics in Medicine},
  year    = {2006},
  volume  = {25},
  number  = {24},
  pages   = {4279--4292},
  doi     = {10.1002/sim.2673}
}

@article{Mallinckrodt2012SensitivityFramework,
  author  = {Mallinckrodt, Craig H. and Lin, Q. and Molenberghs, Geert},
  title   = {A Structured Framework for Assessing Sensitivity to Missing Data Assumptions in Longitudinal Clinical Trials},
  journal = {Pharmaceutical Statistics},
  year    = {2012},
  volume  = {12},
  number  = {1},
  pages   = {1--6},
  doi     = {10.1002/pst.1547}
}

@article{Degtiar2023Generalizability,
  author  = {Degtiar, Irina and Rose, Sherri},
  title   = {A Review of Generalizability and Transportability},
  journal = {Annual Review of Statistics and Its Application},
  year    = {2023},
  volume  = {10},
  number  = {1},
  pages   = {501--524},
  doi     = {10.1146/annurev-statistics-042522-103837}
}

@article{Benda2010ScenarioEvaluation,
  author  = {Benda, Norbert and Branson, Michael and Maurer, Willi and Friede, Tim},
  title   = {Aspects of Modernizing Drug Development Using Clinical Scenario Planning and Evaluation},
  journal = {Drug information journal: DIJ/Drug Information Association},
  year    = {2010},
  volume  = {44},
  number  = {3},
  pages   = {299--315},
  doi     = {10.1177/009286151004400312}
}

@article{Norskov2021AssessmentAssumptions,
  author  = {N{\o}rskov, Anders Kehlet and Lange, Theis and Nielsen, Emil Eik and Gluud, Christian and Winkel, Per and Beyersmann, Jan and de U{\~n}a-{\'A}lvarez, Jacobo and Torri, Valter and Billot, Laurent and Putter, Hein and Wetterslev, J{\o}rn and Thabane, Lehana and Jakobsen, Janus Christian},
  title   = {Assessment of Assumptions of Statistical Analysis Methods in Randomised Clinical Trials: The What and How},
  journal = {BMJ Evidence-Based Medicine},
  year    = {2021},
  volume  = {26},
  number  = {3},
  pages   = {121--126},
  doi     = {10.1136/bmjebm-2019-111268}
}

@article{Marshall2019MIDD,
  author  = {EFPIA MID3 Workgroup; Marshall, SF and Burghaus, R and Cosson, V and Cheung, SY and Chenel, M and DellaPasqua, O and Frey, N and Hamrén, B and Harnisch, L and Ivanow, F and Kerbusch, T and Lippert, J and Milligan, PA and Rohou, S and Staab, A and Steimer, JL and Tornøe, C and Visser, SA},
  title   = {Good Practices in Model-Informed Drug Discovery and Development: Practice, Application, and Documentation},
  journal = {CPT: Pharmacometrics \& Systems Pharmacology},
  year    = {2016},
  volume  = {5},
  number ={3},
  pages   = {93--122},
  doi     = {10.1002/psp4.12049}
}

@article{Burman2005DecisionMaking,
  author  = {Burman, Carl-Fredrik and Hamr{\'e}n, Bengt and Olsson, Per},
  title   = {Modelling and simulation to improve decision-making in clinical development},
  journal = {Pharmaceutical Statistics},
  year    = {2005},
  volume  = {4},
  number  = {1},
  pages   = {47--58},
  doi     = {10.1002/pst.163}
}

@article{Goetghebeur2020CausalQuestions,
  author  = {Goetghebeur, Els and le Cessie, Saskia and De Stavola, Bianca and Moodie, Erica E. M. and Waernbaum, Ingeborg},
  title   = {Formulating causal questions and principled statistical answers},
  journal = {Statistics in Medicine},
  year    = {2020},
  volume  = {39},
  number  = {30},
  pages   = {4922--4948},
  doi     = {10.1002/sim.8741}
}

@article{Nance2024CausalRoadmap,
  author  = {Nance, Nerissa and Petersen, Maya L. and van der Laan, Mark and Balzer, Laura B.},
  title   = {The Causal Roadmap and Simulations to Improve the Rigor and Reproducibility of Real-data Applications},
  journal = {Epidemiology},
  year    = {2024},
  volume  = {35},
  number  = {6},
  pages   = {791--800},
  doi     = {10.1097/EDE.0000000000001773}
}

@article{Lu2025FourTargets,
  author  = {Lu, Haidong and Li, Fan and Lesko, Catherine R. and Fink, David S. and Rudolph, Kara E. and Harhay, Michael O. and Rentsch, Christopher T. and Fiellin, David A. and Gonsalves, Gregg S.},
  title   = {Four targets: an enhanced framework for guiding causal inference from observational data},
  journal = {International Journal of Epidemiology},
  year    = {2025},
  volume  = {54},
  number  = {1},
  doi     = {10.1093/ije/dyaf003}
}

@article{ICHE8R1,
  author = {{ICH}},
  shortauthor      = {{International Council for Harmonization (ICH)}},
  key = {{ICH}},
  title       = {{ICH H}armonised Guideline: General considerations for clinical studies ({E8(R1)})},
  institution = {ICH},
  year        = {2021},
  note        = {Adopted 14 October 2021},
  url         = {https://database.ich.org/sites/default/files/ICH_E8-R1_Guideline_Step4_2021_1006.pdf}
}

@techreport{ICH2024E11A,
  author = {{ICH}},
  shortauthor      = {{International Council for Harmonization (ICH)}},
  key = {{ICH}},
  title       = {{ICH H}armonised Guideline: Pediatric Extrapolation ({E}11{A})},
  institution = {ICH},
  year        = {2024},
  type        = {ICH Guideline},
  note        = {Adopted 21 August 2024},
  url         = {https://database.ich.org/sites/default/files/ICH_E11A_Guideline_Step4_2024_0821.pdf}
}

@article{ICHM15_2026,
  author = {{ICH}},
  shortauthor      = {{International Council for Harmonization (ICH)}},
  key = {{ICH}},
  title       = {{ICH H}armonised Guideline: General Principles for Model-Informed Drug Development ({M}15)},
  institution = {ICH},
  year        = {2026},
  type        = {ICH Guideline},
  note        = {Adopted 29 January 2026},
  url         = {https://database.ich.org/sites/default/files/ICH_M15_Step4_Final_Guideline_2026_0129.pdf}
}

@article{ICHE9,
  author = {{ICH}},
  shortauthor      = {{International Council for Harmonization (ICH)}},
  key = {{ICH}},
  title       = {{ICH H}armonised Tripartite Guideline: Statistical Principles for Clinical Trials ({E9})},
  year        = {1998},
  institution = {ICH},
  note        = {Approved 5 February 1998},
  url         = {https://database.ich.org/sites/default/files/E9_Guideline.pdf}
}

@article{ich2024m15,
  title={{ICH H}armonised Guideline: General Principles for Model-Informed Drug Development ({M15})},
  author = {{ICH}},
  shortauthor={{International Council for Harmonization (ICH)}},
  key = {{ICH}},
  year={2024},
  note={Draft version, Step 2b},
  url = {https://database.ich.org/sites/default/files/ICH_M15_EWG_Step2_DraftGuideline_2024_1031.pdf}
}

@article{ich2019e9,
  title={{ICH H}armonised Guideline: {A}ddendum on {E}stimands and Sensitivity Analysis in Clinical Trials to the Guideline on Statistical Principles for Clinical Trials ({E9(R1)})},
  author = {{ICH}},
  shortauthor={{International Council for Harmonization (ICH)}},
  journal={European Medicines Agency},
  year={2019},
  note        = {Adopted 20 November 2019},
  url         = {https://database.ich.org/sites/default/files/E9-R1_Step4_Guideline_2019_1203.pdf}
}

\end{document}